\documentclass[11pt]{article}%
\usepackage{amsmath}
\usepackage{amssymb}
\usepackage{amsfonts}

\usepackage{cite}
\usepackage{graphicx}
\usepackage{float}
\usepackage{longtable}
\usepackage[utf8]{inputenc}
\usepackage{hyperref}

\usepackage{algorithmicx}
\usepackage[ruled,vlined]{algorithm2e}
\SetAlgoCaptionSeparator{ }

\usepackage{fullpage}

\newcommand{\fref}[1]{Fig.~\ref{#1}}

\newcommand{\tref}[1]{Table~\ref{#1}}
\newcommand{\sref}[1]{Section~\ref{#1}}

\newenvironment{algo}[1][!htbp]
  {
   \begin{algorithm}[#1]%
  }{\end{algorithm}}

\providecommand{\U}[1]{\protect\rule{.1in}{.1in}}
\begin{document}

\title{Adaptive Routing for Quantum Memory Failures in the Quantum Internet}
\author{Laszlo Gyongyosi\thanks{School of Electronics and Computer Science, University of Southampton, Southampton SO17 1BJ, U.K., and Department of Networked Systems and Services, Budapest University of Technology and Economics, 1117 Budapest, Hungary, and MTA-BME Information Systems Research Group, Hungarian Academy of Sciences, 1051 Budapest, Hungary.}
\thanks{Parts of this work were presented in conference proceedings [9].}
\and Sandor Imre\thanks{Department of Networked Systems and Services, Budapest University of Technology and Economics, 1117 Budapest, Hungary.}}\date{}

\maketitle
\begin{abstract}
We define an adaptive routing method for the management of quantum memory failures in the quantum Internet. In the quantum Internet, the entangled quantum states are stored in the local quantum memories of the quantum nodes. A quantum memory failure in a particular quantum node can destroy several entangled connections in the entangled network. A quantum memory failure event makes the immediate and efficient determination of shortest replacement paths an emerging issue in a quantum Internet scenario. The replacement paths omit those nodes that are affected by the quantum memory failure to provide a seamless network transmission. In the proposed solution, the shortest paths are determined by a base-graph, which contains all information about the overlay quantum network. The method provides efficient adaptive routing in quantum memory failure scenarios of the quantum Internet. The results can be straightforwardly applied in practical quantum networks, including long-distance quantum communications.
\end{abstract}
%\begin{keywords}
%quantum Internet - quantum repeater - quantum entanglement - quantum Shannon theory.
%\end{keywords}

\section{Introduction}
\label{sec1}
Entangled quantum networks formulated by quantum repeaters are core elements of the quantum Internet \cite{ref1,ref2,ref3}. In a quantum repeater network, the quantum nodes share quantum entanglement with each other \cite{ref4,ref5,ref6,ref7,ref8,ref9,ref10,ref11,ref12,ref13}, and the shared entangled states are stored in the local quantum memories of the nodes \cite{ref19,ref20,ref21,ref22,ref23,ref24,ref25,ref26,ref27,ref28}. In an entangled network structure, the level of the shared entanglement (number of spanned nodes, i.e., the current level of entanglement swapping can be different, which leads to a heterogeneous multi-level entangled quantum network architecture in general \cite{ref29,ref30,ref31,ref32,ref33,ref34,ref35,ref36,ref37,ref38,ref39}. Because the shared entangled states are stored in the local quantum memories of the quantum nodes, a quantum memory failure in a quantum node can destroy several entangled contacts in the actual shortest main path established between a source and target node \cite{ref40,ref41,ref42,ref43,ref44,ref45,ref46,ref47,ref48}. A quantum memory failure event therefore can have serious consequences in the repeater network, for it leads not only to an immediate requirement for an adaptive routing method that reacts to the dynamic network topology changes, but also to the need for an efficient method for the determination of shortest node-disjoint replacement paths \cite{ref8,ref9,ref10, ref50,ref51,ref52,ref53,ref54,ref55,ref56,ref57,ref58,ref59,ref60,ref61,ref62,ref63}. 

A quantum memory failure event in a quantum node affects a number of entangled connections in the network, and as a corollary, those quantum nodes that have shared entanglement with the given node are also affected by the quantum memory failure. The aim of the node-disjoint \cite{ref14,ref15,ref17} replacement paths is to omit the affected quantum nodes to provide immediate and seamless network transmission. Depending on the number of stored entangled contacts, the quantum nodes can be classified as either standard- or high-degree nodes. Since a memory failure in a high-degree quantum repeater will result in the loss of the highest number entangled connections, a plausible approach would be omitting these nodes from the actual main path. Specifically, this requirement allows us to utilize the omitted high-degree nodes in the replacement paths, based on the assumption that the replacement paths will serve as merely a temporary solution while the re-establishment of the lost entangled contacts is in progress on the main path. 

It is therefore an emerging task to provide seamless transmission in a quantum network during quantum memory failure scenarios, which demands the fast determination of shortest node-disjoint replacement paths between the quantum nodes. Since most of the currently available quantum routing methods \cite{ref1,ref5,ref6,ref7} are based on-, or a variant of Dijkstra's shortest path algorithm \cite{ref49}, the efficiency of these routing approaches is limited. 

Here, we define a dynamic adaptive routing method for the control of quantum memory failures in the quantum Internet. Our method provides a framework for the discovery of shortest node-disjoint replacement paths in the entangled network structure of the quantum Internet. The proposed solution minimizes the number of losable entangled contacts during a quantum memory failure in a main-path node and determines the shortest replacement paths that omit the quantum nodes affected by the memory failure. In particular, all paths are determined in a base-graph \cite{ref8,ref9,ref10} consisting of the maps of the quantum nodes and the entangled connections of the overlay quantum network. The method requires no special devices for the practical implementation, therefore can be applied straightforwardly in quantum networking and in the quantum Internet. Our approach provides a solution for handling quantum memory failures through shortest node-disjoint replacement paths, determined by efficiently adaptive decentralized routing in the quantum Internet.

The novel contributions of our manuscript are as follows:

\begin{enumerate}
\item \textit{We define a dynamic adaptive routing method for the management of quantum memory failures in the quantum Internet.}

\item \textit{The proposed algorithm determines shortest node-disjoint replacement paths in the entangled network structure of the quantum Internet, and minimizes the number of losable entangled contacts in a main-path.}

\item \textit{The shortest node-disjoint replacement paths in the entangled quantum network are determined in a decentralized manner with high computational efficiency.} 
\end{enumerate}

This paper is organized as follows. In \sref{sec2}, we characterize the problem and the impacts of a quantum memory failure scenario in an entangled quantum network. In \sref{sec3}, we discuss the determination of shortest node-disjoint replacement paths and evaluate the complexity of the method. A performance evaluation is included in \sref{sec4}. Finally, \sref{sec5} concludes the paper. Supplementary material is included in the Appendix.

\section{System Model}
\label{sec2}
Let us assume that an overlay entangled quantum network $N$ is given, where the set ${\rm {\mathcal S}}^{{\rm *}} $ of entangled connections is determined in the base-graph $G^{k} $ \cite{ref8} such that $\Pr _{{\rm L}_{l} } \left(x,y\right)\ge \Pr _{{\rm L}_{l} }^{{\rm *}} $, where $\Pr _{{\rm L}_{l} }^{{\rm *}} $ is a threshold probability of an ${\rm L}_{l} $-level entangled connection $E_{h} $. 

Without loss of generality, the level ${\rm L}_{l} $ of an entangled connection is defined as follows. For an ${\rm L}_{l} $-level entangled connection, the hop distance between quantum nodes $x$ and $y$
\begin{equation} \label{1)} 
d\left(x,y\right)_{{\rm L}_{l} } =2^{l-1} , 
\end{equation} 
 with $d\left(x,y\right)_{{\rm L}_{l} } -1$ intermediate nodes between the nodes $x$ and $y$. The probability that an ${\rm L}_{l} $-level entangled connection $E\left(x,y\right)$ exists between nodes $x,y$ is $\Pr _{{\rm L}_{l} } \left(E\left(x,y\right)\right)$, which depends on the actual network.

A repeater node is referred to as a high-degree node, $\tilde{\phi }\left(R_{i} \right)$, if $\deg (\tilde{\phi }\left(R_{i} \right))>\deg '\left(V\right)$, $\tilde{\phi }\left(R_{i} \right)\in G^{k} $, where $\deg '\left(V\right)$ is the threshold degree of node set $V$ determined for the given entangled overlay quantum network $N$. 

In our network model, the high-degree nodes are omitted from the shortest main path ${\rm {\mathcal P}}$. The explanation of this phenomenon is as follows. Since a $\tilde{\phi }\left(R_{i} \right)$ high-degree node stores the highest number of entangled quantum systems in its local quantum memory, a quantum memory failure in $\tilde{\phi }\left(R_{i} \right)$ will result in the loss of the highest number of entangled connections in the quantum network. A rational decision is therefore explained as follows. Use only the standard (non-high-degree nodes) in ${\rm {\mathcal P}}$, since if a quantum memory failure occurs in a standard node $\phi \left(R_{i} \right)\in G^{k} $ of ${\rm {\mathcal P}}$, the high-degree nodes still can participate in a shortest replacement path ${\rm {\mathcal P}}'$. It is because a replacement path is used for only a short time while the re-establishment of the destroyed entangled contacts of $\phi \left(R_{i} \right)$ in ${\rm {\mathcal P}}$ is in progress. As the entangled connections of $\phi \left(R_{i} \right)$ are restored after the memory failure, the transmission continues on the main path ${\rm {\mathcal P}}$. 

Because the main path ${\rm {\mathcal P}}$ omits high-degree nodes, the average length (number of nodes) of the path increases with respect to the shortest path in ${\rm {\mathcal S}}^{{\rm *}} $, but the risk that a high number of entangled connections will vanish during a quantum memory failure is therefore reduced to a minimum. The length of the replacement path ${\rm {\mathcal P}}'$ will be shorter than ${\rm {\mathcal P}}$, for ${\rm {\mathcal P}}'$ can contain the high-degree nodes.

Without loss of generality, in a quantum communication scenario, a quantum path with highest number of entangled connections usually means high communication efficiency and stability \cite{ref1,ref5,ref6,ref7}. It is also the situation for the shortest path ${\rm {\mathcal P}}$ determined by our method, however, these attributes are achieved through the application of an increased number of non high-degree nodes, instead of the use of some high-degree nodes. In the proposed model, a high number of the non-high-degree nodes in the quantum network also yields a high number of entangled connections for the paths; therefore the shortest path will have a high efficiency, reliability and stability.

\begin{center}
\begin{figure*}[!h]
%\vspace{-0.5cm}
\begin{center}
\includegraphics[angle = 0,width=1\linewidth]{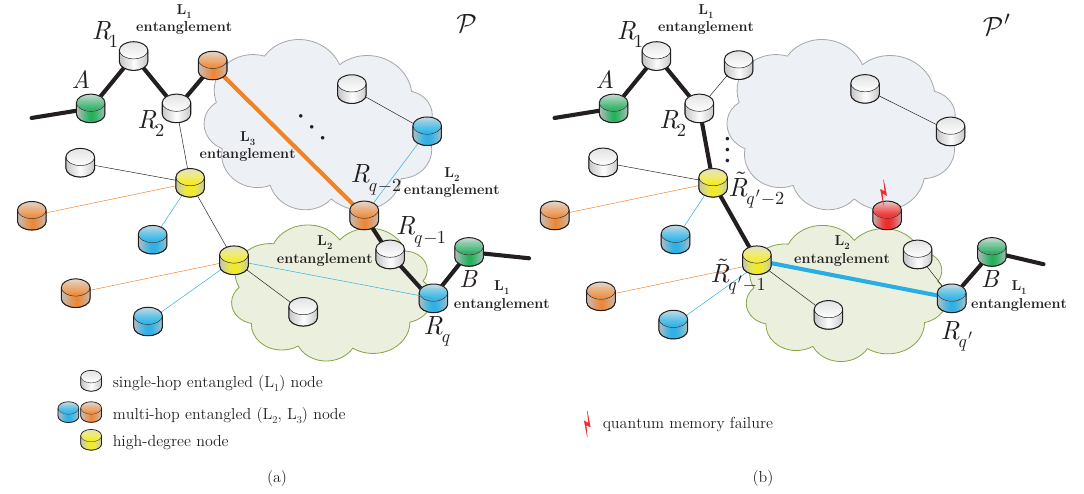}
\caption{A scenario of a quantum memory failure in an entangled overlay quantum network $N$. (a): The main path ${\rm {\mathcal P}}$ consists of a source node $A$ and target node $B$, with several quantum repeater nodes $R_{i} $, $i=1,\ldots ,q$ connected through multi-level entangled connections. The quantum nodes are referred to as single-hop entangled (${\rm L}_{1} $-level, depicted by gray nodes) and multi-hop entangled (${\rm L}_{2} $- and ${\rm L}_{3} $-level, depicted by blue and orange nodes). The actual shortest main path ${\rm {\mathcal P}}$ determined in the base-graph is depicted by the bold lines. The high-degree nodes (depicted by yellow) store the highest number of entangled states in their quantum memories and are not included in ${\rm {\mathcal P}}$. (b): A quantum memory failure in repeater node $R_{q-2} $ (depicted by red) destroys all entangled contacts of the given node. For seamless transmission, a node-disjoint shortest replacement path ${\rm {\mathcal P}}'$ is needed between source node $R_{2} $ and target node $R_{q} $, through high-degree repeater nodes $\tilde{R}_{q'-2} $ and $\tilde{R}_{q'-1} $. Path ${\rm {\mathcal P}}'$ between $\phi \left(R_{2} \right)\in G^{k} $ and $\phi \left(R_{q} \right)\in G^{k} $ is determined by a decentralized routing ${\rm {\mathcal A}}$ in the base-graph $G^{k} $.} 
 \label{fig1}
 \end{center}
\end{figure*}
\end{center}

A quantum memory failure situation in an overlay quantum repeater network $N$ is illustrated \fref{fig1}. The actual shortest path ${\rm {\mathcal P}}$ contains no high-degree repeater nodes, for a memory failure would destroy a high number of entangled contacts in the network (\fref{fig1}(a)). A quantum memory failure in an intermediate (non-high-degree) quantum repeater node $R_{q-2} $ requires the immediate establishment of a replacement shortest path ${\rm {\mathcal P}}'$ in the base-graph. The new shortest path ${\rm {\mathcal P}}'$ contains the high-degree repeater nodes $\tilde{R}_{q'-2} $ and $\tilde{R}_{q'-1} $, for ${\rm {\mathcal P}}'$ is used only while the re-establishment of the entangled contacts of $R_{q-2} $ is not completed (\fref{fig1}(b)).

\subsection{Quantum Memory Failures}

The proposed method utilizes different coefficients for the entangled connections of a shortest main path ${\rm {\mathcal P}}$ and shortest replacement path ${\rm {\mathcal P}}'$, which distinction defines the problem of finding node-disjoint paths of a multi-level entangled quantum network. Particularly, this problem is analogous to a min-sum (minimize the sum of costs of connection paths) problem \cite{ref14,ref15,ref16,ref17} in a multi-cost network, which is an NP-complete problem \cite{ref14,ref16}. Note that similar to the determination of a shortest main path ${\rm {\mathcal P}}$, the replacement path ${\rm {\mathcal P}}'$ is also determined by our decentralized routing ${\rm {\mathcal A}}$ in a base-graph.

\subsubsection{Path Finding}
\label{211}
Here, we summarize the path finding method of \cite{ref8}. The base-graph of an entangled quantum network $N$ is determined as follows. Let $V$ be the set of nodes of the overlay quantum network. Then, let $G^{k} $ be the $k$-dimensional, $n$-sized finite square-lattice base-graph, with position $\phi \left(x\right)$ assigned to an overlay quantum network node $x\in V$, where $\phi :V\to G^{k} $ is a mapping function which achieves the mapping from $V$ onto $G^{k} $ \cite{ref8}. 

Specifically, for two network nodes $x,y\in V$, the L1 metric in $G^{k} $ is $d\left(\phi \left(x\right),\phi \left(y\right)\right)$, $\phi \left(x\right)=\left(j,k\right)$, $\phi \left(y\right)=\left(m,o\right)$ and is defined as
\begin{equation} \label{2)} 
d\left(\left(j,k\right),\left(m,o\right)\right)=\left|m-j\right|+\left|o-k\right|.                                    
\end{equation} 
The $G^{k} $ base-graph contains all entangled contacts of all $x\in V$. The probability that $\phi \left(x\right)$ and $\phi \left(y\right)$ are connected through an ${\rm L}_{l} $-level entanglement in $G^{k} $ is 
\begin{equation} \label{ZEqnNum204871} 
p\left(\phi \left(x\right),\phi \left(y\right)\right)=\frac{d\left(\phi \left(x\right),\phi \left(y\right)\right)^{-k} }{H_{n} } +c_{\phi \left(x\right),\phi \left(y\right)} , 
\end{equation} 
where 
\begin{equation} \label{4)} 
H_{n} =\sum _{z}d\left(\phi \left(x\right),\phi \left(z\right)\right)  
\end{equation} 
is a normalizing term \cite{ref8}, which is taken over all entangled contacts of node $\phi \left(x\right)$ in $G^{k} $, while $c_{\phi \left(x\right),\phi \left(y\right)} $ is a constant defined as
\begin{equation} \label{5)} 
c_{\phi \left(x\right),\phi \left(y\right)} ={{\Pr }_{{\rm L}_{l} }} \left(E\left(x,y\right)\right)-\frac{d\left(\phi \left(x\right),\phi \left(y\right)\right)^{-k} }{H_{n} } , 
\end{equation} 
where $\Pr _{{\rm L}_{l} } \left(E\left(x,y\right)\right)$ is the probability that nodes $x,y\in V$ are connected through an ${\rm L}_{l} $-level entanglement in the overlay quantum network $N$. 

For an ${\rm L}_{l} $-level entanglement between $\phi \left(x\right)$ and $\phi \left(y\right)$, $d\left(\phi \left(x\right),\phi \left(y\right)\right)$ in $G^{k} $ is evaluated as
\begin{equation} \label{6)} 
d\left(\phi \left(x\right),\phi \left(y\right)\right)=2^{l-1} .                                               
\end{equation} 
The routing in the $k$-dimensional base-graph $G^{k} $ is performed via a decentralized algorithm ${\rm {\mathcal A}}$ as follows. After we have determined the base-graph $G^{k} $ of the entangled overlay quantum network $N$, we can apply the L1 metric to find the shortest paths. Since the probability that two arbitrary entangled nodes $\phi \left(x\right),\phi \left(y\right)$ are connected through an ${\rm L}_{l} $-level entanglement is $p\left(\phi \left(x\right),\phi \left(y\right)\right)$ (see \eqref{ZEqnNum204871}), this probability distribution associated with the entangled connectivity in $G^{k} $ allows us to achieve efficient decentralized routing via ${\rm {\mathcal A}}$ in the base-graph. 

Using the L1 distance function, a greedy routing (which always selects a neighbor node closest to the destination node in terms of $G^{k} $ distance function $d$ and does not select the same node twice) can be straightforwardly performed in $G^{k} $ to find the shortest path from any quantum node to any other quantum node, in 
\begin{equation} \label{ZEqnNum549831} 
{\rm {\mathcal O}}\left(\log n\right)^{2}  
\end{equation} 
steps on average, where $n$ is the size of the network of $G^{k} $. 

Note that the nodes know only their local connections (neighbor nodes) and the target position. It also allows us to avoid dead-end nodes (where the routing would stop) by some constraints on the degrees of the nodes, which can be directly satisfied through the settings of the overlay quantum network.

\subsubsection{Shortest Paths at a Quantum Memory Failure}

Using the decentralized routing ${\rm {\mathcal A}}$ (see \sref{211}) in $G'^{k} $, the shortest path with respect to the scaled coefficient $s\left(\zeta \left(\cdot \right)\right)$ can be determined in at most ${\rm {\mathcal O}}\left(\log n\right)^{2} $ steps, as follows.

Using \eqref{ZEqnNum204871}, the term $d\left(\left(\phi \left(x\right),\phi \left(y\right)\right)\right)^{k} $ can be expressed as
\begin{equation} \label{8)} 
d\left(\left(\phi \left(x\right),\phi \left(y\right)\right)\right)^{k} =\frac{p\left(\phi \left(x\right),\phi \left(y\right)\right)}{p\left(\phi \left(x\right),\phi \left(y\right)\right)H_{n} \left(p\left(\phi \left(x\right),\phi \left(y\right)\right)-c_{\phi \left(x\right),\phi \left(y\right)} \right)} , 
\end{equation} 
thus at a given $k$, the $d\left(\left(\phi \left(x\right),\phi \left(y\right)\right)\right)$ distance function between $\phi \left(x\right),\phi \left(y\right)$ is as 
\begin{equation} \label{9)} 
d\left(\left(\phi \left(x\right),\phi \left(y\right)\right)\right)=\sqrt[{k}]{\frac{p\left(\phi \left(x\right),\phi \left(y\right)\right)}{p\left(\phi \left(x\right),\phi \left(y\right)\right)H_{n} \left(p\left(\phi \left(x\right),\phi \left(y\right)\right)-c_{\phi \left(x\right),\phi \left(y\right)} \right)} } . 
\end{equation} 
Using the $\phi \left(x\right),\phi \left(y\right)$ maps of the nodes and the scaled cost $s\left(\zeta \left(E\left(\phi \left(x\right),\phi \left(y\right)\right)\right)\right)\in \left[0,1\right]$, a base-graph $G'^{k} $ is constructed with transformed node positions $\phi '\left(x\right),\phi '\left(y\right)$ as follows.

The aim is then to find $\phi '\left(x\right),\phi '\left(y\right)$ at a given $s\left(\zeta \left(E\left(\phi \left(x\right),\phi \left(y\right)\right)\right)\right)$, via a target distance $d\left(\left(\phi '\left(x\right),\phi '\left(y\right)\right)\right)$.

To determine the $\phi '\left(\cdot \right)$ scaled positions, we first identify $s\left(\zeta \left(E\left(\phi \left(x\right),\phi \left(y\right)\right)\right)\right)$ by $p\left(\phi \left(x\right),\phi \left(y\right)\right)$ in \eqref{ZEqnNum204871} as 
\begin{equation} \label{10)} 
p\left(\phi \left(x\right),\phi \left(y\right)\right)=s\left(\zeta \left(E\left(\phi \left(x\right),\phi \left(y\right)\right)\right)\right),                              
\end{equation} 
from which the target distance $d\left(\left(\phi '\left(x\right),\phi '\left(y\right)\right)\right)$ between $\phi '\left(x\right),\phi '\left(y\right)\in G'^{k} $ is as
\begin{equation} \label{ZEqnNum497976} 
d\left(\left(\phi '\left(x\right),\phi '\left(y\right)\right)\right)=\sqrt[{k}]{\frac{s\left(\zeta \left(E\left(\phi \left(x\right),\phi \left(y\right)\right)\right)\right)}{s\left(\zeta \left(E\left(\phi \left(x\right),\phi \left(y\right)\right)\right)\right)H_{n} \left(s\left(\zeta \left(E\left(\phi \left(x\right),\phi \left(y\right)\right)\right)\right)-c_{\phi \left(x\right),\phi \left(y\right)} \right)} } .   
\end{equation} 
The distance function in \eqref{ZEqnNum497976} is therefore results of the scaled positions $\phi '\left(x\right),\phi '\left(y\right)$ in $G'^{k} $ at a given reference position.

Specifically, to establish a given shortest main path ${\rm {\mathcal P}}$ between a source node $\phi \left(s\right)$ and target node $\phi \left(t\right)$ such that ${\rm {\mathcal P}}$ contains no high-degree repeater node $\tilde{\phi }\left(R_{i} \right)$, a given entangled connection $E_{h} \left(\phi \left(x\right),\phi \left(y\right)\right)$, $E_{h} \in {\rm {\mathcal P}}$ between nodes $\phi \left(x\right),\phi \left(y\right)\in G^{k} $ in the main path ${\rm {\mathcal P}}$ is weighted by the coefficient $\gamma \left(E_{h} \right)$:
\begin{equation} \label{ZEqnNum246623} 
\gamma \left(E_{h} \right)=\frac{\beta \left(\phi \left(x\right)\right)+\beta \left(\phi \left(y\right)\right)}{2} ,                                         
\end{equation} 
where $\beta \left(\phi \left(n\right)\right)$ is defined as a normalization, precisely
\begin{equation} \label{13)} 
\beta \left(\phi \left(n\right)\right)=\frac{\chi \left(\phi \left(n\right)\right)}{\mathop{\max }\limits_{i} \chi \left(\phi \left(i\right)\right)} , 
\end{equation} 
where 
\begin{equation} \label{14)} 
\chi \left(\phi \left(n\right)\right)=\sum _{\phi \left(p\right)\ne \phi \left(q\right)}{\rm s} \frac{\left|{\rm {\mathcal P}}_{\phi \left(n\right)} \left(\phi \left(p\right),\phi \left(q\right)\right)\right|}{\left|{\rm {\mathcal P}}\left(\phi \left(p\right),\phi \left(q\right)\right)\right|} , 
\end{equation} 
where $\left|{\rm {\mathcal P}}_{\phi \left(n\right)} \left(\phi \left(p\right),\phi \left(q\right)\right)\right|$ is the number of shortest paths (of the same minimal length) between nodes $\phi \left(p\right),\phi \left(q\right)\in G^{k} $ and traversing node $\phi \left(n\right)$, and where $\left|{\rm {\mathcal P}}\left(\phi \left(p\right),\phi \left(q\right)\right)\right|$ is the number of shortest paths (of the same minimal length) between nodes $\phi \left(p\right),\phi \left(q\right)\in G^{k} $.  

Let us assume that in an intermediate repeater node $\phi \left(R_{i} \right)$, a quantum memory failure occurs on the main path ${\rm {\mathcal P}}\left(\phi \left(A\right),\phi \left(B\right)\right)$, which destroys all entangled contacts of that node. For seamless communication, an immediate shortest replacement path ${\rm {\mathcal P}}'\left(\phi \left(s\right),\phi \left(r\right)\right)$ has to be established between nodes $\phi \left(s\right),\phi \left(r\right)\in G^{k} $. The replacement shortest path, however, can contain the high-degree nodes $\tilde{\phi }\left(R_{i} \right)$, which have been removed from the main path ${\rm {\mathcal P}}\left(\phi \left(A\right),\phi \left(B\right)\right)$. The replacement shortest path ${\rm {\mathcal P}}'\left(\phi \left(s\right),\phi \left(r\right)\right)$ is aimed to serve as only a temporary path. It replaces the main path ${\rm {\mathcal P}}\left(\phi \left(A\right),\phi \left(B\right)\right)$ while the entangled contacts of $\phi \left(R_{i} \right)$ have not been completely re-established. 

For a given connection $E_{h} $ of the replacement shortest path ${\rm {\mathcal P}}'\left(\phi \left(s\right),\phi \left(r\right)\right)$, the connection coefficient \eqref{ZEqnNum246623} is redefined as 
\begin{equation} \label{ZEqnNum107944} 
\tau \left(E_{h} \right)=\frac{\gamma \left(E_{h} \right)}{\mathop{\max }\limits_{i} \gamma \left(E_{i} \right)} ,                                                
\end{equation} 
which provides a normalized metric for $E_{h} $ in ${\rm {\mathcal P}}'\left(\phi \left(s\right),\phi \left(r\right)\right)$. 

Using coefficients \eqref{ZEqnNum246623} and \eqref{ZEqnNum107944}, the optimization problem formulates a minimization, without loss of generality as
\begin{equation} \label{ZEqnNum433702} 
\Phi \left(C\right)=\min \left(\sum _{k\in K}\sum _{h\in E}\gamma \left(E_{h} \right)C_{k,h} +\sum _{k\in K}\sum _{h\in E}\tau \left(E_{h} \right)Z_{k,h}     \right), 
\end{equation} 
where $K$ is the set of users, $k$ is the index of a given user $U_{k} $, $E$ is a set of entangled (directed) connections, $h$ is the index of a given connection $E_{h} $, $C_{k,h} $ is a variable that equals 1 if the entangled connection $E_{h} $ is used by the main path ${\rm {\mathcal P}}$ associated with user $U_{k} $ (0 otherwise), and $Z_{k,h} $ is a variable that equals 1 if the entangled connection $E_{h} $ is used by the replacement path ${\rm {\mathcal P}}'$ associated with user $U_{k} $.

The optimization in \eqref{ZEqnNum433702} is therefore a minimization of the overall cost coefficient of the flows by means of main path ${\rm {\mathcal P}}$ and replacement path ${\rm {\mathcal P}}'$, subject to some constraints. These constraints are described next.

Particularly, for the main path ${\rm {\mathcal P}}$, a flow conservation rule \cite{ref14,ref16} leads to precisely
\begin{equation} \label{17)} 
{\rm \triangle }\left(C_{k,h} \right)=\sum _{{\rm {\mathcal X}}_{h,j} }C_{k,h}  -\sum _{{\rm {\mathcal X}}_{h,i} }C_{k,h}  ,                                            
\end{equation} 
where
\begin{equation} \label{ZEqnNum523994} 
\begin{array}{l} {{\rm {\mathcal X}}_{h,j} =h\in \left\{h:E_{h} \left(\phi \left(x_{w} \right)_{U_{k} } ,\phi \left(x_{j} \right)_{U_{k} } \right)\in E;\right. } \\ {{\rm \; \; \; \; \; \; \; \; \; \; \; \; \; }\phi \left(x_{j} \right)_{U_{k} } \in V;} \\ {\left. {\rm \; \; \; \; \; \; \; \; \; \; \; \; \; }\phi \left(x_{j} \right)_{U_{k} } \ne \phi \left(x_{w} \right)_{U_{k} } \right\},} \end{array} 
\end{equation} 
where $\phi \left(x_{w} \right)_{U_{k} } \in G^{k} $ is a node associated with the demand of user $k\in K$, where $E_{h} \left(\phi \left(x_{w} \right)_{U_{k} } ,\phi \left(x_{j} \right)_{U_{k} } \right)$ is an egress connection incident associated with node $\phi \left(x_{w} \right)_{U_{k} } \in G^{k} $, where $E_{h} \left(\phi \left(x_{i} \right)_{U_{k} } ,\phi \left(x_{w} \right)_{U_{k} } \right)$ is an ingress connection incident associated with node $\phi \left(x_{w} \right)_{U_{k} } \in G^{k} $, and where
\begin{equation} \label{ZEqnNum869067} 
\begin{array}{l} {{\rm {\mathcal X}}_{h,i} =h\in \left\{h:E_{h} \left(\phi \left(x_{i} \right)_{U_{k} } ,\phi \left(x_{w} \right)_{U_{k} } \right)\in E;\right. } \\ {{\rm \; \; \; \; \; \; \; \; \; \; \; \; \; }\phi \left(x_{i} \right)_{U_{k} } \in V;} \\ {\left. {\rm \; \; \; \; \; \; \; \; \; \; \; \; \; }\phi \left(x_{i} \right)_{U_{k} } \ne \phi \left(x_{w} \right)_{U_{k} } \right\}.} \end{array} 
\end{equation} 
Using \eqref{ZEqnNum523994} and \eqref{ZEqnNum869067}, ${\rm \triangle }\left(C_{k,h} \right)$ can be evaluated as
\begin{equation} \label{20)} 
{\rm \triangle }\left(C_{k,h} \right)=\left\{\begin{array}{l} {1,{\rm \; }\text{if}{\rm \; }\phi \left(x_{w} \right)_{U_{k} } =\phi \left(A\right)_{U_{k} } } \\ {-1,{\rm \; }\text{if}{\rm \; }\phi \left(x_{w} \right)_{U_{k} } =\phi \left(B\right)_{U_{k} } } \\ {0,{\rm \; otherwise\; \; \; \; \; \; \; \; \; \; \; \; \; \; \; \; }} \end{array}\right. . 
\end{equation} 
Similarly, for the replacement path ${\rm {\mathcal P}}'$, the following constraint is defined via \eqref{ZEqnNum523994} and \eqref{ZEqnNum869067}:
\begin{equation} \label{21)} 
{\rm \triangle }\left(Z_{k,h} \right)=\sum _{{\rm {\mathcal X}}_{h,j} }Z_{k,h}  -\sum _{{\rm {\mathcal X}}_{h,i} }Z_{k,h}  =\left\{\begin{array}{l} {1,{\rm \; }\text{if}{\rm \; }\phi \left(x_{w} \right)_{U_{k} } =\phi \left(A\right)_{U_{k} } } \\ {-1,{\rm \; }\text{if}{\rm \; }\phi \left(x_{w} \right)_{U_{k} } =\phi \left(B\right)_{U_{k} } } \\ {0,{\rm \; otherwise\; \; \; \; \; \; \; \; \; \; \; \; \; \; \; \; }} \end{array}\right. . 
\end{equation} 
For a given entangled connection $E_{h} $, the following relation states that the requested number of entangled states of a particular fidelity $F$ on the main path ${\rm {\mathcal P}}$ and replacement path ${\rm {\mathcal P}}'$ is bounded as:
\begin{equation} \label{22)} 
\varphi \left(E_{h} \right)=\sum _{k\in K}\left(C_{k,h} Q^{\left(F\right)} \left(U_{k} \right)+Z_{k,h} Q^{\left(F\right)} \left(U_{k} \right)\right) \le Q^{\left(F\right)} \left(E_{h} \right),                   
\end{equation} 
where $Q^{\left(F\right)} \left(E_{h} \right)$ refers to the number of $d$-dimensional maximally entangled states per second of a particular fidelity $F$ available through the entangled connection $E_{h} $ (entanglement throughput), and where $Q^{\left(F\right)} \left(U_{k} \right)$ is the demand of $U_{k} $ associated with the entangled connection $E_{h} $ with respect to the number of $d$-dimensional maximally entangled states per second of a particular fidelity $F$.

Because ${\rm {\mathcal P}}$ and ${\rm {\mathcal P}}'$ are node-disjoint paths in our model, the following constraint holds for a given $E_{h} $ and $U_{k} $:
\begin{equation} \label{23)} 
C_{k,h} +Z_{k,h} \le 1.                                                  
\end{equation} 

\section{Shortest Replacement Paths in the Entangled Network}
\label{sec3}
Here we discuss a heuristic to solve \eqref{ZEqnNum433702} via the determination of a set of $z$ node-disjoint replacement paths ${\rm S}^{\left(z\right)} \left({\rm {\mathcal P}}'\right)=\left\{{\rm {\mathcal P}}'_{1} ,\ldots ,{\rm {\mathcal P}}'_{z} \right\}$ in the base-graph $G^{k} $ of an entangled quantum repeater network $N$. The algorithm focuses on a given demand $\rho $ of a user $U_{k} $, with a source node $\phi \left(s\right)_{U_{k} ,\rho } \in G^{k} $ and target node $\phi \left(t\right)_{U_{k} ,\rho } \in G^{k} $. 

Let $\delta \left(\cdot \right)$ identify the cost function of the algorithm such that if a given connection $E_{h} $ belongs to the main path ${\rm {\mathcal P}}$, then $\delta \left(E_{h} \right)=\gamma \left(E_{h} \right)$ \eqref{ZEqnNum246623}, whereas if $E_{h} $ belongs to a replacement path ${\rm {\mathcal P}}'$, then $\delta \left(E_{h} \right)=\tau \left(E_{h} \right)$ \eqref{ZEqnNum107944}. Thus,
\begin{equation} \label{24)} 
\delta \left(E_{h} \right)=\left\{\begin{array}{l} {\gamma \left(E_{h} \right), \text{if}{\rm \; }E_{h} \in {\rm {\mathcal P}}} \\ {\tau \left(E_{h} \right), \text{if}{\rm \; }E_{h} \in {\rm {\mathcal P}}'} \end{array}\right. . 
\end{equation} 
Utilizing the notations of the frameworks KPA \cite{ref16,ref17} and KPI \cite{ref14,ref15}, some preliminaries for our scheme are as follows. Let set ${\rm S}^{\left(j-1\right)} \left({\rm {\mathcal P}}'\right)=\left\{{\rm {\mathcal P}}'_{1} ,\ldots ,{\rm {\mathcal P}}'_{j-1} \right\}$ refer to the previously discovered $j-1$ node-disjoint paths, where ${\rm {\mathcal P}}'_{1} $ refers to the main path, i.e., ${\rm {\mathcal P}}'_{1} ={\rm {\mathcal P}}$. Specifically, for each node-disjoint path ${\rm {\mathcal P}}_{p} $, $p=1,\ldots ,z$, a cost matrix $C_{U_{k} ,\rho }^{\left(p\right)} $ is defined, which is a matrix of connection costs $\delta ^{\left(p\right)} \left(E_{h} \right)$, where $\delta ^{\left(p\right)} \left(E_{h} \right)$ is an auxiliary cost of entangled connection $E_{h} $ of the $p$-th path. The $C_{U_{k} ,\rho }^{\left(p\right)} $ matrix is used to calculate the concurring path cost such that the cost of the concurring entangled connections is increasing by a given value.

Particularly, for a given set ${\rm S}^{\left(j-1\right)} \left({\rm {\mathcal P}}'\right)$ of already discovered $j-1$ paths, a $j$-th path identifies a current (candidate) path ${\rm {\mathcal P}}'_{j} $ to be discovered. A given entangled connection $E_{h} $ of a path ${\rm {\mathcal P}}'_{i} $ is identified as a prohibited entangled connection ${\rm {\mathcal F}}^{\left({\rm {\mathcal P}}'_{i} \right)} \left(E_{h} \right)$ with respect to ${\rm {\mathcal P}}'_{i} $ if $E_{h} $ is incident to any transit quantum node of path ${\rm {\mathcal P}}'_{i} $ \cite{ref14,ref16}. 

A given $E_{h} $ is referred to as a concurring entangled connection ${\rm {\mathcal C}}^{\left({\rm {\mathcal P}}'_{j} \right)} \left(E_{h} \right)$ with respect to a given path ${\rm {\mathcal P}}'_{j} $ if $E_{h} $ is incident to any common transit quantum node of ${\rm {\mathcal P}}'_{j} $ also used by any other of the paths from the set ${\rm S}^{\left(j-1\right)} \left({\rm {\mathcal P}}'\right)$ of the previously discovered $j-1$ paths. Without loss of generality, the initial matrix $C_{U_{k} ,\rho }^{\left(p\right)} $ provides the initial path cost $c^{\left(j\right)} =\sum _{E_{h} \in {\rm {\mathcal P}}'_{j} }\delta _{c}^{\left(j\right)} \left(E_{h} \right) $ for a given path ${\rm {\mathcal P}}'_{j} $ to increase the cost of each concurring entangled connection of ${\rm {\mathcal P}}'_{j} $, where $\delta _{c}^{\left(j\right)} \left(E_{h} \right)$ is an initial cost of entangled connection $E_{h} $ and where $E_{h} \in {\rm {\mathcal P}}'_{j} $ is an entangled connection on path ${\rm {\mathcal P}}'_{j} $. 

Let $M_{U_{k} ,\rho }^{\delta ^{\left(j\right)} \left(E_{h} \right)} $ be the matrix of replacement path coefficients, with $\delta ^{\left(j\right)} \left(E_{h} \right)$ for all entangled connections of a current path ${\rm {\mathcal P}}'_{j} $ associated with a user $U_{k} $. For a given matrix $M_{U_{k} ,\rho }^{\delta ^{\left(j\right)} \left(E_{h} \right)} $, let $\Omega ^{\left(j\right)} \left(\delta \right)$ refer to the total cost of a path ${\rm {\mathcal P}}'_{j} $. 

For a current path index $j$, let $M_{U_{k} ,\rho }^{\left(\zeta \right)} $ refer to a matrix of coefficients $\zeta \left(E_{h} \right)=\delta ^{\left(j\right)} \left(E_{h} \right)$ of entangled connection $E_{h} $, where $\zeta \left(E_{h} \right)$ is an auxiliary cost of $E_{h} $. 

The steps of the method are summarized in Algorithm 1. The algorithm determines $z$ node-disjoint paths for a demand of a user. The main path ${\rm {\mathcal P}}$ is identified first and followed by the $z-1$ replacement paths. For a given $j$-th path, the cost of any prohibited entangled connection is increased by the cost of all previously discovered $j-1$ paths of demand for which the given entangled connection is prohibited \cite{ref14,ref16}. Traversing the prohibited entangled connections with respect to a given $j$-th path therefore results in an increased coefficient. The cost of concurring entangled connections increases if a given $j$-th path has common entangled connections with the $j-1$ paths.

 \setcounter{algocf}{0}
\begin{algo}
  \DontPrintSemicolon
\caption{Node-Disjoint Replacement Paths in an Entangled Network}

\textbf{Step 1}. Let $\delta ^{\left(j\right)} \left(E_{h} \right)$ be the cost coefficient of $E_{h} $ on a $j$-th path $j=1\ldots z$, where $j$ indexes a current path. Let $j=1$ for the main path ${\rm {\mathcal P}}'_{j=1} ={\rm {\mathcal P}}$, and for all entangled connections $E_{h} $ of $N$, let $\zeta \left(E_{h} \right)=\delta ^{\left(j\right)} \left(E_{h} \right)$, where $\zeta \left(E_{h} \right)$ is an auxiliary cost of $E_{h} $.

\textbf{Step 2}. For $j>1$, let us assume that a set ${\rm S}^{\left(j-1\right)} \left({\rm {\mathcal P}}'\right)$ of $j-1$ node-disjoint shortest paths is already discovered: ${\rm S}^{\left(j-1\right)} \left({\rm {\mathcal P}}'\right)=\left\{{\rm {\mathcal P}}'_{1} ,\ldots ,{\rm {\mathcal P}}'_{j-1} \right\}$. Let $i$ refer to a path from ${\rm S}^{\left(j-1\right)} \left({\rm {\mathcal P}}'\right)$. For each already discovered node-disjoint path ${\rm {\mathcal P}}'_{i} $ from ${\rm S}^{\left(j-1\right)} \left({\rm {\mathcal P}}'\right)$ and for each ${\rm {\mathcal F}}^{\left({\rm {\mathcal P}}'_{i} \right)} \left(E_{h} \right)$, increase the cost of the entangled connection $\zeta \left(E_{h} \right)$ by $\Omega ^{\left(i\right)} $, where $\Omega ^{\left(i\right)} $ is the total cost of ${\rm {\mathcal P}}'_{i} $ in $M_{U_{k} ,\rho }^{\delta ^{\left(i\right)} \left(E_{h} \right)} $ and where $M_{U_{k} ,\rho }^{\delta ^{\left(i\right)} \left(E_{h} \right)} $ is a matrix such that $\delta ^{\left(i\right)} \left(E_{h} \right)$ is the cost of an entangled connection $E_{h} $ on the $i$-th path. Thus,
\[\zeta \left(E_{h} \right)=\zeta \left(E_{h} \right)+\Omega ^{\left(i\right)} ,\] 
where $\Omega ^{\left(i\right)} $ is the path cost of ${\rm {\mathcal P}}'_{i} $ (a sum of $\delta ^{\left(i\right)} \left(E_{h} \right)$ for a traversed entangled connection $E_{h} $): 
\[\Omega ^{\left(i\right)} =\sum _{E_{h} \in {\rm {\mathcal P}}'_{i} }\delta ^{\left(i\right)} \left(E_{h} \right),i=1,\ldots ,j-1 ,\] 
where $E_{h} \in {\rm {\mathcal P}}'_{i} $ refers to $E_{h} $ on path ${\rm {\mathcal P}}'_{i} $.    

\textbf{Step 3}. Define $M_{U_{k} ,\rho }^{\left(\zeta \right)} =M_{U_{k} ,\rho }^{\delta ^{\left(j\right)} \left(E_{h} \right)} $, where $M_{U_{k} ,\rho }^{\left(\zeta \right)} $ is a matrix of coefficients $\zeta \left(E_{h} \right)$ of entangled connection $E_{h} $ initialized in Step 1. Using the $k$-dimensional $n$-size base-graph $G'^{k} $, determine the $j$-th node-disjoint path ${\rm {\mathcal P}}'_{j} $ using the scaled coefficient $s\left(\zeta \left(E_{h} \right)\right)\in \left[0,1\right]$ for a given $M_{U_{k} ,\rho }^{\left(\zeta \right)} $.

\textbf{Step 4}. If ${\rm {\mathcal P}}'_{j} $ is not node-disjoint with the paths of ${\rm S}^{\left(j-1\right)} \left({\rm {\mathcal P}}'\right)$, then increase the costs $\delta ^{\left(1\right)} \left(E_{h} \right),\ldots ,\delta ^{\left(k\right)} \left(E_{h} \right)$ of each concurring entangled connection ${\rm {\mathcal C}}^{\left({\rm {\mathcal P}}'_{j} \right)} \left(E_{h} \right)$ of ${\rm {\mathcal P}}'_{j} $ by the path cost $c^{\left(j\right)} \in C_{U_{k} ,\rho }^{\left(j\right)} $ of ${\rm {\mathcal P}}'_{j} $ with cost matrix $C_{U_{k} ,\rho }^{\left(j\right)} $. Thus, the cost of entangled connection $E_{h} $ of the $p$-th path is 
\[\delta ^{\left(p\right)} \left(E_{h} \right)=\delta ^{\left(p\right)} \left(E_{h} \right)+c^{\left(j\right)} , p=1,\ldots ,z,\] 
where path cost $c^{\left(j\right)} $ is 
\[c^{\left(j\right)} =\sum _{E_{h} \in {\rm {\mathcal P}}'_{j} }\delta _{c}^{\left(j\right)} \left(E_{h} \right) .\] 
Remove the discovered $j-1$ paths from ${\rm S}^{\left(j-1\right)} \left({\rm {\mathcal P}}'\right)$, and register the concurring entangled connection ${\rm {\mathcal C}}^{\left({\rm {\mathcal P}}'_{j} \right)} \left(E_{h} \right)$ via variable $\kappa :=\kappa +1$, where $\kappa $ is initialized as $\kappa =1$. If $\kappa >\partial $, where $\partial $ is the maximum allowable number of concurrences, then terminate the procedure; otherwise, repeat the process from Step 1 with $j=1$.

\end{algo}

 \setcounter{algocf}{0}
\begin{algo}
  \DontPrintSemicolon
\caption{Node-Disjoint Replacement Paths in an Entangled Network (cont.)}

\textbf{Step 5}. If ${\rm {\mathcal P}}'_{j} $ is node-disjoint with the paths of ${\rm S}^{\left(j-1\right)} \left({\rm {\mathcal P}}'\right)$, then increase $j$: $j:=j+1$. If $j>z$, stop the process and return ${\rm S}^{\left(z\right)} \left({\rm {\mathcal P}}'\right)$; otherwise, go to Step 1 with the current $j$.

\textbf{Step 6}. For a given main path ${\rm {\mathcal P}}'_{1} ={\rm {\mathcal P}}$, output the $\Psi $ total path cost of $z-1$ node-disjoint replacement paths between $\phi \left(s\right)_{U_{k} ,\rho } $ and $\phi \left(t\right)_{U_{k} ,\rho } $ of $\rho $:
\[\Psi \left({\rm S}^{\left(z-1\right)} \left({\rm {\mathcal P}}'\right)\right)=\sum _{p=2}^{z}\sum _{E_{h} \in {\rm {\mathcal P}}'_{p} }\delta ^{\left(p\right)} \left(E_{h} \right)  ,\] 
where ${\rm S}^{\left(z-1\right)} \left({\rm {\mathcal P}}'\right)=\left\{{\rm {\mathcal P}}'_{2} ,\ldots ,{\rm {\mathcal P}}'_{z} \right\}$, $E_{h} $ is an entangled connection on path ${\rm {\mathcal P}}_{p} ^{\lefteqn{\mkern0mu\lower4pt\hbox{$ {'}  $}}\phantom{{'} }} $, and $\delta ^{\left(p\right)} \left(E_{h} \right)$ is the cost of the entangled connection $E_{h} $ of the $p$-th path. 

\end{algo}

\subsection{Discussion}

A brief description of the algorithm follows. 

In Step 1, some initialization steps are made and an actual shortest main path ${\rm {\mathcal P}}'_{1} ={\rm {\mathcal P}}$ is determined for the next calculations. 

In Step 2, some steps for the next ($j$-th) node-disjoint path (candidate path) are performed. In particular, the coefficients of the prohibited entangled connections that are traversed by the actual main path are increased by a given quantity. This step aims to avoid a situation in which the establishment of the next node-disjoint path of a given user fails. Some calculations are performed for the $j$-th path using the already determined set of $j-1$ node-disjoint paths ${\rm S}^{\left(j-1\right)} \left({\rm {\mathcal P}}'\right)$. The cost of any prohibited entangled connection is increased by the total cost of a given path ${\rm {\mathcal P}}'_{i} $.  

In Step 3, the $j$-th disjoint (shortest) path ${\rm {\mathcal P}}'_{j} $ is determined by the decentralized algorithm ${\rm {\mathcal A}}$ in the base-graph $G'^{k} $, which contains the scaled $\phi '\left(x\right),\phi '\left(y\right)\in G'^{k} $ of the nodes of $\phi \left(x\right),\phi \left(y\right)\in G^{k} $. The base-graph $G'^{k} $ is evaluated from $G^{k} $, and it contains the $\phi '$ scaled maps of the nodes of the entangled overlay quantum network $N$ and the scaled  coefficients of the entangled connection, $s\left(\zeta \left(E_{h} \right)\right)$ using $M_{U_{k} ,\rho }^{\left(\zeta \right)} $. 

In $G'^{k} $, a given contact between two nodes $\phi '\left(x\right),\phi '\left(y\right)$ is characterized by the scaled coefficient $s\left(\zeta \left(E\left(\phi \left(x\right),\phi \left(y\right)\right)\right)\right)\in \left[0,1\right]$, where $\zeta \left(E\left(\phi \left(x\right),\phi \left(y\right)\right)\right)$ is the cost of an entangled connection $E_{h} $ in $G^{k} $. Particularly, $G'^{k} $ is constructed such that the distribution of the scaled coefficients follows an inverse $k$-power distribution, and the decentralized routing scheme ${\rm {\mathcal A}}$ can determine the shortest path in $G'^{k} $ in at most \eqref{ZEqnNum549831} steps \cite{ref8}. 

Step 4 deals with the situation when the $j$-th path ${\rm {\mathcal P}}'_{j} $ is not node-disjoint with the paths of ${\rm S}^{\left(j-1\right)} \left({\rm {\mathcal P}}'\right)$. If more than one already discovered path from ${\rm S}^{\left(j-1\right)} \left({\rm {\mathcal P}}'\right)$ traverses a given entangled connection, then the cost of the concurring entangled connection is increased by the total cost of path ${\rm {\mathcal P}}'_{j} $. Step 4 is completed by the incrementing and checking of a $\kappa $ concurrence counter.

Step 5 handles the case when a $j$-th path ${\rm {\mathcal P}}'_{j} $ is node-disjoint with the paths of ${\rm S}^{\left(j-1\right)} \left({\rm {\mathcal P}}'\right)$. In this situation, $j$ is incremented: $j:=j+1$; if $j>z$, the iteration stops and returns the $z-1$ node-disjoint shortest replacement paths ${\rm S}^{\left(z-1\right)} \left({\rm {\mathcal P}}'\right)=\left\{{\rm {\mathcal P}}'_{2} ,\ldots ,{\rm {\mathcal P}}'_{z} \right\}$ for a given main path ${\rm {\mathcal P}}'_{1} ={\rm {\mathcal P}}'$; otherwise, the algorithm jumps back to Step 1 with the actual value of $j$.

In Step 6, after set ${\rm S}^{\left(z-1\right)} \left({\rm {\mathcal P}}'\right)$ is determined, the $\Psi $ total cost of the $z-1$ replacement paths of demand $\rho $ of $U_{k} $ is precisely as follows:
\begin{equation} \label{25)} 
\Psi \left({\rm S}^{\left(z-1\right)} \left({\rm {\mathcal P}}'\right)\right)=\sum _{p=2}^{z}\sum _{E_{h} \in {\rm {\mathcal P}}'_{p} }\delta ^{\left(p\right)} \left(E_{h} \right)  ,                                   
\end{equation} 
where $E_{h} $ is an entangled connection on path ${\rm {\mathcal P}}_{p} ^{\lefteqn{\mkern0mu\lower4pt\hbox{$ {'}  $}}\phantom{{'} }} $. The $\Psi $ total cost for the set ${\rm S}^{\left(z\right)} \left({\rm {\mathcal P}}'\right)$ of the $z$ node-disjoint paths is therefore
\begin{equation} \label{26)} 
\Psi \left({\rm S}^{\left(z\right)} \left({\rm {\mathcal P}}'\right)\right)=\sum _{p=1}^{z}\sum _{E_{h} \in {\rm {\mathcal P}}'_{p} }\delta ^{\left(p\right)} \left(E_{h} \right)  .                                    
\end{equation} 
 
\begin{center}
\begin{figure*}[!h]
%\vspace{-0.5cm}
\begin{center}
\includegraphics[angle = 0,width=1\linewidth]{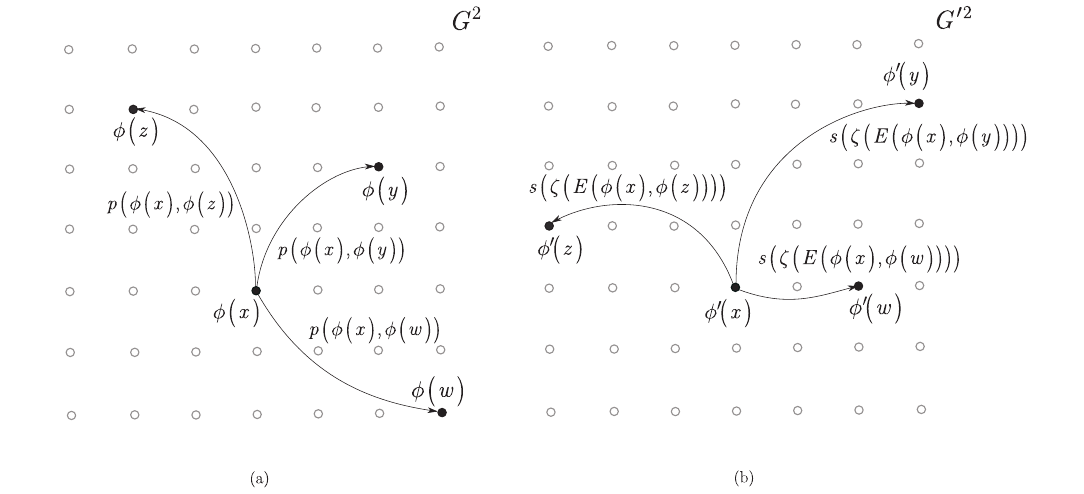}
\caption{A $k=2$-dimensional base-graph $G^{2} $ of the overlay quantum repeater network $N$. (a): A reference source node $\phi \left(x\right)$ has entangled connections with $\phi \left(y\right)$, $\phi \left(z\right)$, and $\phi \left(w\right)$. Each entangled connection is characterized by a given probability $p\left(\cdot \right)$ that depends on the level of entanglement. (b): Determination of a node-disjoint path ${\rm {\mathcal P}}'_{j} $ between a reference source node $\phi '\left(x\right)$ and the scaled positions $\phi '\left(y\right)$, $\phi '\left(z\right)$, and $\phi '\left(w\right)$ for a given $M_{U_{k} ,\rho }^{\left(\zeta \right)} $ in base-graph $G'^{2} $, where $s\left(\zeta \left(E\left(\cdot \right)\right)\right)\in \left[0,1\right]$ is a scaled cost of the entangled connection.} 
 \label{fig2}
 \end{center}
\end{figure*}
\end{center}

The base-graphs $G^{k} $ and $G'^{k} $ of a given overlay quantum repeater network $N$ are illustrated in \fref{fig2}.   

\subsection{Computational Complexity}

In particular, each of the shortest paths is determined by a decentralized routing algorithm ${\rm {\mathcal A}}$ in a $k$-dimensional $n$-size base-graph, therefore the overall complexity the proposed method is bounded from above by 
\begin{equation} \label{ZEqnNum734745} 
{\rm {\mathcal O}}\left(\partial \log n\right)^{2} ,                                                    
\end{equation} 
for a given maximum allowable number $\partial $ of concurring entangled connections.

\subsection{Practical Implementations}

An experimental implementation of the proposed method in a stationary quantum node can be achieved by standard photonics devices, quantum memories, optical cavities and other fundamental physical devices \cite{ref11,ref12,ref13,ref28,ref38,ref39,ref40,ref41,ref42,ref43,ref44} required for practical quantum network communications \cite{ref46,ref47,ref48,ref50,ref51,ref52,ref53,ref54,ref55,ref56,ref57,ref58,ref59,ref60}. The quantum transmission and the auxiliary classical communications between the nodes can be realized via standard links (e.g., optical fibers, wireless quantum channels, free-space optical channels, etc) via the integration of the fundamental quantum transmission protocols of quantum networks \cite{ref1}. 

\section{Performance Evaluation}
\label{sec4}
In this section, we compare the performance of our scheme with the KPA, KPI, and the $k$-successively shortest link-disjoint paths (KSP) \cite{ref15} algorithms.  

The KPA and KPI algorithms are based on Dijkstra's shortest path \cite{ref49} algorithm. As follows, at a particular $\partial $, the computational complexity of these schemes is ${\rm {\mathcal O}}\left(\partial n^{2} \right)$ \cite{ref14,ref15,ref17}.

The worst-case complexity of the KSP algorithm can be evaluated in function of the number $z$ of disjoint paths, as ${\rm {\mathcal O}}\left(zn\log n\right)$ \cite{ref15}.

\begin{center}
\begin{figure*}[!h]
%\vspace{-0.5cm}
\begin{center}
\includegraphics[angle = 0,width=1\linewidth]{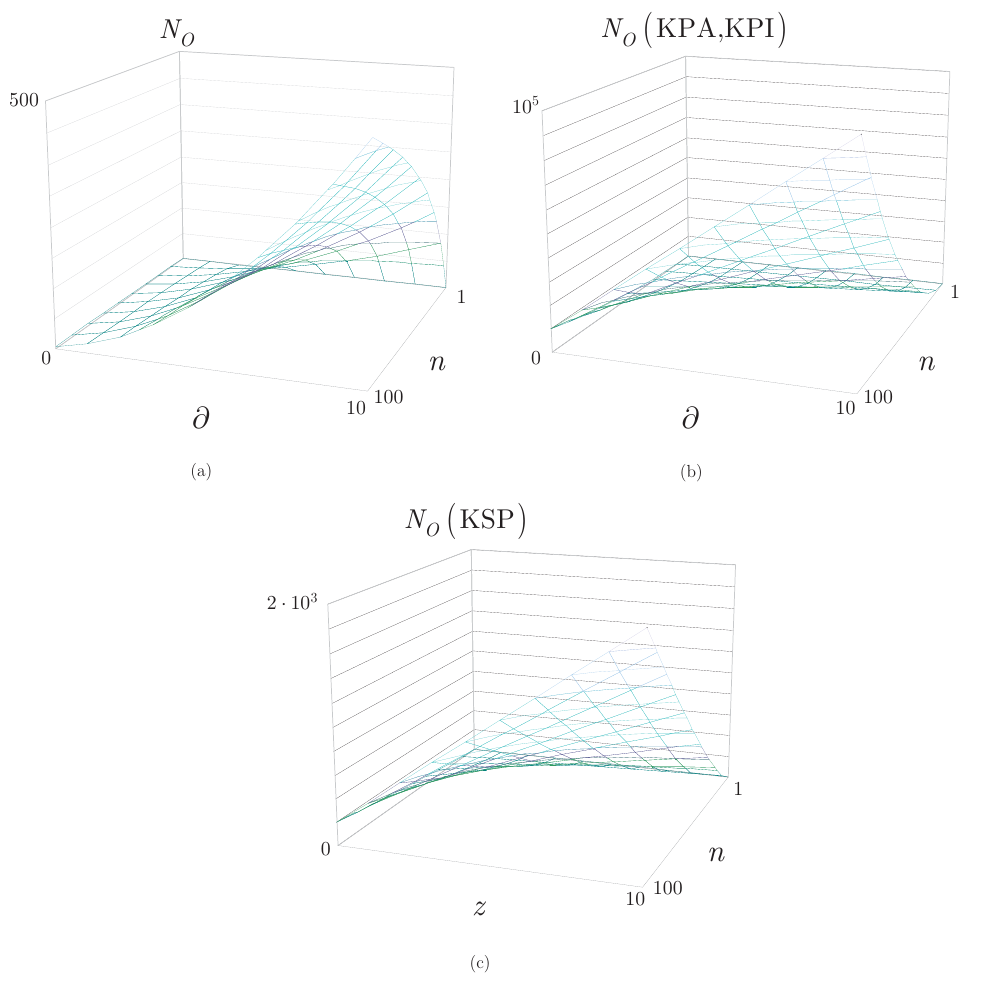}
\caption{(a): The computational complexity ($N_{O} $ refers to the number of operations) of the proposed method in function of $\partial $ and $n$, $\partial \in \left[0,10\right]$, $n\in \left[0,100\right]$ (b): The computational complexity of the KPI and KPA methods in function of $\partial $ and $n$, $\partial \in \left[0,10\right]$, $n\in \left[0,100\right]$. (c): The computational complexity of the KSP method in function of $z$ and $n$, $z\in \left[0,10\right]$, $n\in \left[0,100\right]$.} 
 \label{fig3}
 \end{center}
\end{figure*}
\end{center}

Fig. 3. illustrates the performance comparisons, and $N_{O} $ refers to the number of operations. The performance of our scheme is depicted \fref{fig3}(a). In \fref{fig2}(b) the performance of the KPA, KPI algorithms is depicted. In \fref{fig3}(c) the performance of the KSP algorithm is shown. 

For the analyzed parameter ranges, in our algorithm $N_{O} $ is maximized as $N_{O} =400$, while for the compared methods the resulting quantities are $N_{O} \left({\rm KPA,KPI}\right)=10^{5} $, and $N_{O} \left({\rm KSP}\right)=2\cdot 10^{3} $, respectively. 

As a conclusion, in comparison with the KPA, KPI and KSP methods, our solution provides a moderate complexity solution to determine the connection-disjoint paths in the quantum network, for an arbitrary number of $n$ and $\partial $.

\section{Conclusions}
\label{sec5}
We defined an adaptive routing scheme for the handling of quantum memory failures in the entangled quantum networks of the quantum Internet. The main path contains no high-degree repeater nodes, for in a case of quantum memory failure, these quantum nodes result in the highest number of lost entangled contacts in the network. The method finds the shortest node-disjoint replacement paths between source and target quantum nodes. The replacement paths serve as temporary paths until all destroyed entangled contacts are completely re-established between the repeater nodes. All path-searching phases are performed on a base-graph of the overlay quantum network to provide an efficient computational solution. The scheme can be directly applied in practical quantum networks, including long-distance quantum communications. 

%
%\section*{Statements}
%\subsection*{Ethics statement}
%This work did not involve any active collection of human data.
%\subsection*{Data accessibility statement}
%This work does not have any experimental data.
%\subsection*{Competing financial interests statement}
%We have no competing financial interests.
%\subsection*{Competing interests statement}
%We have no competing interests.
%\subsection*{Funding}
%No relevant funding. 
%\subsection*{Authors’ contributions}
%L.GY. designed the protocol and wrote the manuscript. L.GY. and S.I. analyzed the results. All authors reviewed the manuscript.
%\subsection*{Original Article Statement}
%a. This paper is a completely novel and completely independent submission, has no any connection with any previously submitted papers.
%b. This manuscript is the authors' original work and has not been published nor has it been submitted simultaneously elsewhere. 
%c. All authors have checked the manuscript and have agreed to the submission. 

\section*{Acknowledgements}
This work was partially supported by the National Research Development and Innovation Office of Hungary (Project No. 2017-1.2.1-NKP-2017-00001), by the Hungarian Scientific Research Fund - OTKA K-112125 and in part by the BME Artificial Intelligence FIKP grant of EMMI (BME FIKP-MI/SC).

\newpage
%\onecolumn
\appendix
\setcounter{table}{0}
\setcounter{figure}{0}
\setcounter{equation}{0}
\setcounter{algocf}{0}
\renewcommand{\thetable}{\Alph{section}.\arabic{table}}
\renewcommand{\thefigure}{\Alph{section}.\arabic{figure}}
\renewcommand{\theequation}{\Alph{section}.\arabic{equation}}
\renewcommand{\thealgocf}{\Alph{section}.\arabic{algocf}}

\setlength{\arrayrulewidth}{0.1mm}
\setlength{\tabcolsep}{5pt}
\renewcommand{\arraystretch}{1.5}
\section{Appendix}
\subsection{Abbreviations}
\begin{description}
\item[NP] Nondeterministic Polynomial time
\item[KPA] K-Penalty
\item[KPI] K-Penalty with Initial cost matrix
\item[KSP] K-Successively shortest link disjoint Paths

\end{description}

\subsection{Notations}
The notations of the manuscript are summarized in  \tref{tab2}.
\begin{center}
\begin{longtable}{||l|p{4.5in}||}
\caption{Summary of notations.}
\label{tab2}
\endfirsthead
\endhead
\hline
\textit{Notation} & \textit{Description} \\ \hline
L1 & Manhattan distance (L1 metric). \\ \hline 
$l$  & Level of entanglement.  \\ \hline 
$F$ & Fidelity of entanglement.  \\ \hline 
${\rm L}_{l} $ & An $l$-level entangled connection. For an ${\rm L}_{l} $ entangled connection, the hop-distance is $2^{l-1} $. \\ \hline 
$d\left(x,y\right)_{{\rm L}_{l} } $ & Hop-distance of an $l$-level entangled connection between nodes $x$ and $y$.  \\ \hline 
${\rm L}_{1} $ & ${\rm L}_{1} $-level (direct) entanglement,  $d\left(x,y\right)_{{\rm L}_{1} } =2^{0} =1$. \\ \hline 
${\rm L}_{2} $ & ${\rm L}_{2} $-level entanglement, $d\left(x,y\right)_{{\rm L}_{2} } =2^{1} =2$. \\ \hline 
${\rm L}_{3} $ & ${\rm L}_{3} $-level entanglement, $d\left(x,y\right)_{{\rm L}_{3} } =2^{2} =4$. \\ \hline 
$E\left(x,y\right)$ & An edge between quantum nodes $x$ and $y$, refers to an ${\rm L}_{l} $-level entangled connection. \\ \hline 
$\Pr _{{\rm L}_{l} } \left(E\left(x,y\right)\right)$ & Probability of existence of an entangled connection $E\left(x,y\right)$, $0<\Pr _{{\rm L}_{l} } \left(E\left(x,y\right)\right)\le 1$. \\ \hline 
$N$ & Overlay quantum network, $N=\left(V,E\right)$, where $V$ is the set of nodes, $E$ is the set of edges. \\ \hline 
$V$ & Set of nodes of $N$. \\ \hline 
$E$ & Set of edges of $N$. \\ \hline 
$G^{k} $ & An $n$-size, $k$-dimensional base-graph. \\ \hline 
$n$ & Size of base-graph $G^{k} $. \\ \hline 
$k$  & Dimension of base-graph $G^{k} $. \\ \hline 
$A$ & Transmitter node, $A\in V$. \\ \hline 
$B$ & Receiver node, $B\in V$. \\ \hline 
$R_{i} $ & A repeater node in $V$, $R_{i} \in V$. \\ \hline 
$E_{j} $ & Identifies an ${\rm L}_{l} $-level entanglement, $l=1,\ldots ,r$, between quantum nodes $x_{j} $ and $y_{j} $. \\ \hline 
$E=\left\{E_{j} \right\}$ & Let $E=\left\{E_{j} \right\}$, $j=1,\ldots ,m$ refer to a set of edges between the nodes of $V$. \\ \hline 
$\phi \left(x\right)$ & Position assigned to an overlay quantum network node $x\in V$ in a $k$-dimensional, $n$-sized finite square-lattice base-graph $G^{k} $. \\ \hline 
$\phi :V\to G^{k} $ & Mapping function that achieves the mapping from $V$ onto $G^{k} $. \\ \hline 
$d\left(\phi \left(x\right),\phi \left(y\right)\right)$ & L1 distance between $\phi \left(x\right)$ and $\phi \left(y\right)$ in $G^{k} $. For   $\phi \left(x\right)=\left(j,k\right)$, $\phi \left(y\right)=\left(m,o\right)$ evaluated as\newline $d\left(\left(j,k\right),\left(m,o\right)\right)=\left|m-j\right|+\left|o-k\right|$. \\ \hline 
$p\left(\phi \left(x\right),\phi \left(y\right)\right)$ & The probability that $\phi \left(x\right)$ and $\phi \left(y\right)$ are connected through an ${\rm L}_{l} $-level entanglement in $G^{k} $. \\ \hline 
$H_{n} $ & Normalizing term, defined as $H_{n} =\sum _{z}d\left(\phi \left(x\right),\phi \left(z\right)\right) $. \\ \hline 
$c_{\phi \left(x\right),\phi \left(y\right)} $ & Constant, defined as\newline $c_{\phi \left(x\right),\phi \left(y\right)} =\Pr _{{\rm L}_{l} } \left(E\left(x,y\right)\right)-\frac{d\left(\phi \left(x\right),\phi \left(y\right)\right)^{-k} }{H_{n} } ,$\newline where $\Pr _{{\rm L}_{l} } \left(E\left(x,y\right)\right)$ is the probability that nodes $x,y\in V$ are connected through an ${\rm L}_{l} $-level entanglement in the overlay quantum network $N$. \\ \hline 
$\Pr \left(\left. E\right|\phi \right)$ & Conditional probability between the $\phi \left(\cdot \right)$ configuration of positions of the quantum nodes in $G^{k} $ and the set $E$ of the $m$ edges of the overlay network $V$. \\ \hline 
$\Pr \left(\left. \phi \right|E\right)$ & Posteriori distribution of configuration $\phi $ at a given set $E$. \\ \hline 
$\Pr \left(\phi \right)$ & Candidate distribution. \\ \hline 
$\tilde{\phi }\left(R_{i} \right)$ & A high-degree quantum repeater node, \newline $\deg (\tilde{\phi }\left(R_{i} \right))>\deg '\left(V\right)$,\newline where $\deg '\left(V\right)$ is the threshold degree of node set. \\ \hline 
${\rm {\mathcal P}}$ & A shortest main path. \\ \hline 
${\rm {\mathcal P}}'$ & A shortest replacement path. \\ \hline 
${\rm {\mathcal S}}^{{\rm *}} $ & A set of entangled connections, such that $\Pr _{{\rm L}_{l} } \left(x,y\right)\ge \Pr _{{\rm L}_{l} }^{{\rm *}} $, where $\Pr _{{\rm L}_{l} }^{{\rm *}} $ is a threshold probability of an ${\rm L}_{l} $-level entangled connection $E_{h} $. \\ \hline 
${\rm {\mathcal A}}$ & Decentralized algorithm ${\rm {\mathcal A}}$ in the $k$-dimensional $n$-sized base-graph $G^{k} $. \\ \hline 
$D\left(G^{k} \right)$ & Diameter of $G^{k} $. Refers to the maximum value of the shortest path (total number of edges on a path) between any pair of mapped nodes in $G^{k} $. \\ \hline 
$D\left({\rm {\mathcal A}}\right)$ & Minimal number of steps required by ${\rm {\mathcal A}}$ to find the shortest path. \\ \hline 
$\gamma \left(E_{h} \right)$ & Weighting coefficient of the main path ${\rm {\mathcal P}}$. \\ \hline 
$\beta \left(\phi \left(n\right)\right)$ & Normalization term, evaluated as\newline $\beta \left(\phi \left(n\right)\right)=\frac{\chi \left(\phi \left(n\right)\right)}{\mathop{\max }\limits_{i} \chi \left(\phi \left(i\right)\right)} ,$\newline where \newline $\chi \left(\phi \left(n\right)\right)=\sum _{\phi \left(p\right)\ne \phi \left(q\right)}{\rm s} \frac{\left|{\rm {\mathcal P}}_{\phi \left(n\right)} \left(\phi \left(p\right),\phi \left(q\right)\right)\right|}{\left|{\rm {\mathcal P}}\left(\phi \left(p\right),\phi \left(q\right)\right)\right|} ,$                               \newline where $\left|{\rm {\mathcal P}}_{\phi \left(n\right)} \left(\phi \left(p\right),\phi \left(q\right)\right)\right|$ is the number of shortest paths (of the same minimal length) between $\phi \left(p\right),\phi \left(q\right)\in G^{k} $ and traversing node $\phi \left(n\right)$, while $\left|{\rm {\mathcal P}}\left(\phi \left(p\right),\phi \left(q\right)\right)\right|$ is the number of shortest paths (of the same minimal length) between nodes $\phi \left(p\right),\phi \left(q\right)\in G^{k} $. \\ \hline 
$\tau \left(E_{h} \right)$ & A normalization term for $E_{h} $ in ${\rm {\mathcal P}}'\left(\phi \left(s\right),\phi \left(r\right)\right)$, as \newline $\tau \left(E_{h} \right)=\frac{\gamma \left(E_{h} \right)}{\mathop{\max }\limits_{i} \gamma \left(E_{i} \right)} $. \\ \hline 
$\Phi \left(C\right)$ & Subject of a minimization in an optimization.  \\ \hline 
$K$ & Set of users. \\ \hline 
$k$ & Index of a given user $U_{k} $. \\ \hline 
$E$ & A set of entangled connections. \\ \hline 
$h$ & Index of a given entangled connection $E_{h} $. \\ \hline 
$C_{k,h} $ & A variable, equals 1 if the entangled connection $E_{h} $ is used by the main path ${\rm {\mathcal P}}$ associated with user $U_{k} $, and 0 otherwise. \\ \hline 
$Z_{k,h} $ & A variable, equals 1 if the entangled connection $E_{h} $ is used by the replacement path ${\rm {\mathcal P}}'$ associated with user $U_{k} $, and 0 otherwise. \\ \hline 
${\rm \triangle }\left(C_{k,h} \right)$ & A flow conservation rule parameter. \\ \hline 
$\phi \left(x_{w} \right)_{U_{k} } $ & A node associated with the demand of user $U_{k} $, $\phi \left(x_{w} \right)_{U_{k} } \in G^{k} $. \\ \hline 
$E_{h} \left(\phi \left(x_{w} \right)_{U_{k} } ,\phi \left(x_{j} \right)_{U_{k} } \right)$ & An egress entangled connection incident associated with node $\phi \left(x_{w} \right)_{U_{k} } \in G^{k} $. \\ \hline 
$E_{h} \left(\phi \left(x_{i} \right)_{U_{k} } ,\phi \left(x_{w} \right)_{U_{k} } \right)$ & An ingress entangled connection incident associated with node $\phi \left(x_{w} \right)_{U_{k} } \in G^{k} $. \\ \hline 
$\varphi \left(E_{h} \right)$ & A variable, the requested number of entangled states of a particular fidelity $F$ on the main path ${\rm {\mathcal P}}$ and replacement path ${\rm {\mathcal P}}'$ is bounded by the entangled connection by this quantity. \\ \hline 
$Q^{\left(F\right)} \left(U_{k} \right)$ & The demand of $U_{k} $ associated with the entangled connection $E_{h} $ with respect to the number of $d$-dimensional maximally entangled states per second of a particular fidelity $F$. \\ \hline 
$\delta \left(\cdot \right)$ & Variable, identifies the cost function of the algorithm such that if a given entangled connection $E_{h} $ belongs to the main path ${\rm {\mathcal P}}$, then $\delta \left(E_{h} \right)=\gamma \left(E_{h} \right)$, whereas if $E_{h} $ belongs to a replacement path ${\rm {\mathcal P}}'$, then $\delta \left(E_{h} \right)=\tau \left(E_{h} \right)$. \\ \hline 
${\rm S}^{\left(j-1\right)} \left({\rm {\mathcal P}}'\right)$ & A set ${\rm S}^{\left(j-1\right)} \left({\rm {\mathcal P}}'\right)=\left\{{\rm {\mathcal P}}'_{1} ,\ldots ,{\rm {\mathcal P}}'_{j-1} \right\}$, refers to the previously discovered $j-1$ node-disjoint paths, where ${\rm {\mathcal P}}'_{1} $ refers to the main path, i.e., ${\rm {\mathcal P}}'_{1} ={\rm {\mathcal P}}$. \\ \hline 
$C_{U_{k} ,\rho }^{\left(p\right)} $ & Matrix of path costs, used to calculate the concurring path cost such that the cost of the concurring entangled connections is increasing by a given value. \\ \hline 
$\delta ^{\left(p\right)} \left(E_{h} \right)$ & An auxiliary cost of entangled connection $E_{h} $ of the $p$-th path. \\ \hline 
${\rm {\mathcal P}}'_{j} $ & A $j$-th path, identifies a current (candidate) path ${\rm {\mathcal P}}'_{j} $ to be discovered. \\ \hline 
${\rm {\mathcal F}}^{\left({\rm {\mathcal P}}'_{i} \right)} \left(E_{h} \right)$ & Prohibited entangled connection. A given entangled connection $E_{h} $ of a path ${\rm {\mathcal P}}'_{i} $ is identified as a ${\rm {\mathcal F}}^{\left({\rm {\mathcal P}}'_{i} \right)} \left(E_{h} \right)$ with respect to ${\rm {\mathcal P}}'_{i} $ if $E_{h} $ is incident to any transit quantum node of path ${\rm {\mathcal P}}'_{i} $. \\ \hline 
${\rm {\mathcal C}}^{\left({\rm {\mathcal P}}'_{j} \right)} \left(E_{h} \right)$ & Concurring entangled connection. A given $E_{h} $ is referred to as a ${\rm {\mathcal C}}^{\left({\rm {\mathcal P}}'_{j} \right)} \left(E_{h} \right)$ with respect to a given path ${\rm {\mathcal P}}'_{j} $ if $E_{h} $ is incident to any common transit quantum node of ${\rm {\mathcal P}}'_{j} $ also used by any other of the paths from the set ${\rm S}^{\left(j-1\right)} \left({\rm {\mathcal P}}'\right)$ of the previously discovered $j-1$ paths. \\ \hline 
$\delta _{c}^{\left(j\right)} \left(E_{h} \right)$ & An initial cost of entangled connection $E_{h} $. \\ \hline 
$E_{h} \in {\rm {\mathcal P}}'_{j} $ & An entangled connection on path ${\rm {\mathcal P}}'_{j} $. \\ \hline 
$M_{U_{k} ,\rho }^{\delta ^{\left(j\right)} \left(E_{h} \right)} $ & Matrix of replacement path coefficients, with $\delta ^{\left(j\right)} \left(E_{h} \right)$ for all entangled connections of a current path ${\rm {\mathcal P}}'_{j} $ associated with a user $U_{k} $. \\ \hline 
$\Omega ^{\left(j\right)} \left(\delta \right)$ & The total cost of a path ${\rm {\mathcal P}}'_{j} $.  \\ \hline 
$M_{U_{k} ,\rho }^{\left(\zeta \right)} $ & Matrix of coefficients $\zeta \left(E_{h} \right)=\delta ^{\left(j\right)} \left(E_{h} \right)$ of entangled connection $E_{h} $, where $\zeta \left(E_{h} \right)$ is an auxiliary cost of $E_{h} $ for a current path index $j$. \\ \hline 
$z$ & Number of node-disjoint paths for a demand of a user.  \\ \hline 
$\Psi $ & Total path cost of $z-1$ node-disjoint replacement paths between $\phi \left(s\right)_{U_{k} ,\rho } $ and $\phi \left(t\right)_{U_{k} ,\rho } $ of $\rho $,\newline $\Psi \left({\rm S}^{\left(z-1\right)} \left({\rm {\mathcal P}}'\right)\right)=\sum _{p=2}^{z}\sum _{E_{h} \in {\rm {\mathcal P}}'_{p} }\delta ^{\left(p\right)} \left(E_{h} \right)  $,\newline where ${\rm S}^{\left(z-1\right)} \left({\rm {\mathcal P}}'\right)=\left\{{\rm {\mathcal P}}'_{2} ,\ldots ,{\rm {\mathcal P}}'_{z} \right\}$, $E_{h} $ is an entangled connection on path ${\rm {\mathcal P}}_{p} ^{\lefteqn{\mkern0mu\lower4pt\hbox{$ {'}  $}}\phantom{{'} }} $, and $\delta ^{\left(p\right)} \left(E_{h} \right)$ is the cost of the entangled connection $E_{h} $ of the $p$-th path.  \\ \hline 
${\rm S}^{\left(z-1\right)} \left({\rm {\mathcal P}}'\right)$ & The total cost of the $z-1$ replacement paths of demand $\rho $ of $U_{k} $,\newline $\Psi \left({\rm S}^{\left(z-1\right)} \left({\rm {\mathcal P}}'\right)\right)=\sum _{p=2}^{z}\sum _{E_{h} \in {\rm {\mathcal P}}'_{p} }\delta ^{\left(p\right)} \left(E_{h} \right)  $,\newline where $E_{h} $ is an entangled connection on path ${\rm {\mathcal P}}_{p} ^{\lefteqn{\mkern0mu\lower4pt\hbox{$ {'}  $}}\phantom{{'} }} $. \\ \hline 
$\Psi \left({\rm S}^{\left(z\right)} \left({\rm {\mathcal P}}'\right)\right)$ & The $\Psi $ total cost for the set ${\rm S}^{\left(z\right)} \left({\rm {\mathcal P}}'\right)$ of the $z$ node-disjoint paths, \newline $\Psi \left({\rm S}^{\left(z\right)} \left({\rm {\mathcal P}}'\right)\right)=\sum _{p=1}^{z}\sum _{E_{h} \in {\rm {\mathcal P}}'_{p} }\delta ^{\left(p\right)} \left(E_{h} \right)  $. \\ \hline
 \end{longtable}
\end{center}
\end{document}